\input harvmac
\input epsf

%%%%%%%%%%%%%%%%%%%%%%%%%%%%%%%%%%%%%%%%%
\def\eql{~=~}

\def\al{\alpha}

\def\coeff#1#2{\relax{\textstyle {#1 \over #2}}\displaystyle}
\def\half{{1 \over 2}}

 \def\cM{{\cal M}}
\def\cN{{\cal N}} \def\cO{{\cal O}}
\def\cP{{\cal P}} 
 
 \def\cV{{\cal V}}
\def\ie{{\it i.e.}}
\def\bfone{\relax{\rm 1\kern-.35em 1}}
\def\IC{\relax\,\hbox{$\inbar\kern-.3em{\rm C}$}}
\def\ID{\relax{\rm I\kern-.18em D}}
\def\IF{\relax{\rm I\kern-.18em F}}
\def\IH{\relax{\rm I\kern-.18em H}}
\def\II{\relax{\rm I\kern-.17em I}}
\def\IN{\relax{\rm I\kern-.18em N}}
\def\IP{\relax{\rm I\kern-.18em P}}
\def\IQ{\relax\,\hbox{$\inbar\kern-.3em{\rm Q}$}}
\def\us#1{\underline{#1}}
\def\IR{\relax{\rm I\kern-.18em R}}
\font\cmss=cmss10 \font\cmsss=cmss10 at 7pt
\def\ZZ{\relax\ifmmode\mathchoice
{\hbox{\cmss Z\kern-.4em Z}}{\hbox{\cmss Z\kern-.4em Z}}
{\lower.9pt\hbox{\cmsss Z\kern-.4em Z}}
{\lower1.2pt\hbox{\cmsss Z\kern-.4em Z}}\else{\cmss Z\kern-.4em
Z}\fi}
\def\eop{\mathop{e}^{\!\circ}{}}
\def\gop{\mathop{g}^{\!\circ}{}}
%
%
%%%%%%%%%%%%%%%%%%%%%%%%%%%%%%%%%%%%%%%%%
%%% Referencing
%%%%%%%%%%%%%%%%%%%%%%%%%%%%%%%%%%%%%%%%%
\def\nihil#1{{\it #1}}
\def\eprt#1{{\tt #1}}
%%%%%%%%%%%%%%%%%%%%%%%%%%%%%%%%%%%%%%%%%
\def\nup#1({Nucl.\ Phys.\ $\us {B#1}$\ (}
\def\plt#1({Phys.\ Lett.\ $\us  {#1B}$\ (}
\def\cmp#1({Comm.\ Math.\ Phys.\ $\us  {#1}$\ (}
\def\prp#1({Phys.\ Rep.\ $\us  {#1}$\ (}
\def\prl#1({Phys.\ Rev.\ Lett.\ $\us  {#1}$\ (}
\def\prv#1({Phys.\ Rev.\ $\us  {#1}$\ (}
\def\mpl#1({Mod.\ Phys.\ Let.\ $\us  {A#1}$\ (}
\def\ijmp#1({Int.\ J.\ Mod.\ Phys.\ $\us{A#1}$\ (}
\def\jag#1({Jour.\ Alg.\ Geom.\ $\us {#1}$\ (}

%%%%%%%%%%%%%%%%%%%%%%%%%%%%%%%%%%%%%%%%%
% References
%%%%%%%%%%%%%%%%%%%%%%%%%%%%%%%%%%%%%%%%%

\lref\PWcrpt{K.\ Pilch and N.P.\ Warner, \nihil{A New Supersymmetric
Compactification of Chiral IIB Supergravity},
\eprt{hep-th/0002192}.}
\lref\KPW{A.\ Khavaev, K.\ Pilch and N.P.\ Warner, \nihil{New Vacua of
Gauged  ${\cal N}=8$ Supergravity in Five Dimensions},
\eprt{hep-th/9812038}.}
\lref\GRW{M.\ G\"unaydin, L.J.\ Romans and N.P.\ Warner,
\nihil{Gauged $N=8$ Supergravity in Five Dimensions,}
Phys.~Lett.~{\bf 154B} (1985) 268; \nihil{Compact and Non-Compact
Gauged Supergravity Theories in Five Dimensions,}
\nup{272} (1986) 598.}
\lref\PPvN{M.~Pernici, K.~Pilch and P. van Nieuwenhuizen,
\nihil{Gauged $N=8$, $D = 5$ Supergravity,} \nup{259} (1985) 460.}
\lref\LJR{L.J.\ Romans, \nihil{New Compactifications of Chiral $N=2$,
$d=10$ Supergravity,}  \plt{153} (1985) 392.}
\lref\RLMS{R.~G. Leigh and M.~J. Strassler, \nihil{Exactly Marginal
Operators and
Duality in Four-Dimensional $N=1$ Supersymmetric Gauge Theory,}
\nup{447} (1995) 95; \eprt{hep-th/9503121}}
\lref\FGPWa{D.~Z. Freedman, S.~S. Gubser, K.~Pilch, and N.~P. Warner,
\nihil{Renormalization Group Flows from Holography---Supersymmetry
and a c-Theorem,} CERN-TH-99-86, \eprt{hep-th/9904017} }
\lref\JSIIB{J.H.\ Schwarz, \nihil{Covariant Field Equations of
Chiral $N=2$, $D=10$ Supergravity,} CALT-68-1016,
\nup{226} (1983) 269.}
\lref\PvNW{P.~van~Nieuwenhuizen and N.P. Warner, \nihil{New
Compactifications of Ten-Dimensional and Eleven-Dimensional
Supergravity on Manifolds which are not Direct Products} \cmp{99}
(1985) 141.}
\lref\MetAns{B.\ de Wit and H.\ Nicolai, \nihil{On the Relation
Between $d=4$ and $d=11$ Supergravity,} Nucl.~Phys.~{\bf B243}
(1984) 91; \hfil \break
B.\ de Wit, H.\ Nicolai and N.P.\ Warner,
\nihil{The Embedding of Gauged $N=8$ Supergravity into $d=11$
Supergravity,}
Nucl.~Phys.~{\bf B255} (1985) 29.}
\lref\adscftrev{
O.~Aharony, S.~S.~Gubser, J.~Maldacena, H.~Ooguri and Y.~Oz,
\nihil{Large N Field Theories, String Theory and Gravity,}
\eprt{hep-th/9905111}. }
\lref\DiLithium{B.\ de Wit, H.\ Nicolai,  \nihil{A New $SO(7)$
Invariant Solution of $d = 11$ Supergravity,}
Phys.~Lett.~{\bf 148B} (1984) 60.}
\lref\NWstrings{N.P.\ Warner, \nihil{Renormalization Group Flows from
Five-Dimensional Supergravity,} talk presented at Strings `99, Potsdam,
Germany, 19--25 Jul 1999; \eprt{hep-th/9911240}}
\lref\JSIIB{J.H.~Schwarz, \nihil{Covariant Field Equations of Chiral,
$N=2$ $D=10$ Supergravity,} Nucl.~Phys.~{\bf B226}  (1983) 269.}
\lref\NSVZ{
V.~Novikov, M.~A.\ Shifman, A.~I.\ Vainshtein, V.~Zakharov, {\it Exact
Gell-Mann-Low Function of Supersymmetric Yang-Mills Theories from
Instanton Calculus,} Nucl. Phys. {\bf B229} (1983) 381.}
\lref\UCAK{U.\ Chattopadhyay and A.\ Karlhede,
\nihil{Consistent Truncation Of Kaluza-Klein Theories,}
 Phys.\ Lett.\ {\bf 139B} 279 (1984).}
\lref\FGPWb{D.~Z. Freedman, S.~S. Gubser, K.~Pilch, and N.~P. Warner,
{\it Continuous Distribution of D3-branes and Gauged Supergravity,}
\eprt{hep-th/9906194}. }
\lref\JMalda{J.~Maldacena, \nihil{The Large $N$ Limit of Superconformal
Field Theories and Supergravity,}, Adv.~Theor. Math. Phys.~{\bf 2}
(1998) 231 \eprt{hep-th/9711200}.}
\lref\WitHolOne{E.\ Witten, \nihil{Anti-de Sitter space and
holography,} Adv. Theor.  Math.  Phys. {\bf 2} (1998) 253,
\eprt{hep-th/9802150}.}
\lref\SGIKAP{S.S.\ Gubser, I.R.\ Klebanov, A.M.\ Polyakov,
\nihil{Gauge Theory  Correlators from Non-Critical String Theory,}
Phys.~Lett.~{\bf B428}  (1998) 105, \eprt{hep-th/9802109}.}
\lref\PolStr{J.~Polchinski and M.~J.~Strassler,
{\it The String Dual of a Confining Four-Dimensional Gauge Theory,}
\eprt{hep-th/0003136}.}
\lref\JPP{C.~V.~Johnson, A.~W.~Peet and J.~Polchinski,
{\it  Gauge Theory and the Excision of Repulson Singularities,}
Phys. Rev.\ {\bf D61} (2000) 086001, \eprt{hep-th/9911161}.}
\lref\Trunseven{H. Nastase, D. Vaman and P. van Nieuwenhuizen,
{\it Consistent Nonlinear KK Reduction of 11d Supergravity on
$AdS_7\times S_4$ and Self-Duality in Odd Dimensions,}
Phys. Lett. {\bf B469} (1999) 96, \eprt{hep-th/9905075};
{\it Consistency of the $AdS_7\times S_4$ Reduction and the
Origin of  Self-Duality in Odd Dimensions,} \eprt{hep-th/9911238}.}
\lref\NastVam{H. Nastase and D. Vaman, {\it On the Nonlinear KK
Reductions
on Spheres of Supergravity Theories,} \eprt{hep-th/0002028}.}
\lref\Cetal{M.~Cvetic {\it et al.},
\nihil{Embedding AdS black holes in ten and eleven dimensions,}
Nucl.\ Phys.\  {\bf B558} (1999) 96, \eprt{hep-th/9903214}.}
\lref\CGLPx{
M.~Cvetic, S.~S.~Gubser, H.~Lu and C.~N.~Pope,
\nihil{Symmetric potentials of gauged supergravities in diverse 
dimensions and  Coulomb branch of gauge theories},
\eprt{hep-th/9909121}.}
\lref\LPTx{
H.~Lu, C.~N.~Pope and T.~A.~Tran,
\nihil{Five-dimensional N = 4, SU(2) x U(1) gauged supergravity from 
type IIB},
\eprt{hep-th/9909203}.}
\lref\CLPSx{
M.~Cvetic, H.~Lu, C.~N.~Pope and A.~Sadrzadeh,
\nihil{Consistency of Kaluza-Klein sphere reductions of symmetric potentials},
\eprt{hep-th/0002056}.}
\lref\CLP{M.~Cveti\v{c}, H.~L\"u and C.N.~Pope, \nihil{Geometry of
the Embedding of Scalar Manifolds in $D=11$ and $D=10$,  }
\eprt{hep-th/0002099}.}
\lref\Popeetala{
M. Cvetic, H. Lu, C. N. Pope, A. Sadrzadeh and T. A. Tran,
\nihil{Consistent SO(6) reduction of type IIB supergravity on
$S_5$}, \eprt{hep-th/0003103}.}
\lref\SGsing{S.~Gubser, \nihil{Curvature Singularities:  The Good,
The Bad, and the Naked,} PUPT-1916,
\eprt{hep-th/0002160}.}
\lref\GPPZ{L. Girardello, M. Petrini, M. Porrati and A.
Zaffaroni \nihil{The supergravity dual of N = 1 super Yang-Mills theory ,}
Nucl. Phys. {\bf B569} (2000) 451, \eprt{hep-th/9909047}.}
\lref\GPPZold{L. Girardello, M. Petrini, M. Porrati and A.
Zaffaroni \nihil{Novel local CFT and exact results on 
perturbations of N = 4 super  Yang-Mills from AdS dynamics},
JHEP {\bf 12} (1998) 022, \eprt{hep-th/9810126}.}
\lref\DistZam{J. Distler and F. Zamora, \nihil{Non-supersymmetric 
conformal field theories from stable anti-de Sitter  spaces},
Adv. Theor. Math. Phys. {\bf 2} (1999) 1405, \eprt{hep-th/9810206};
\nihil{Chiral symmetry breaking in the AdS/CFT correspondence},
\eprt{hep-th/9911040}.}
\lref\CastPes{L. Castellani and I. Pesando,
\nihil{The Complete Superspace Action of Chiral D = 10, N=2
                  Supergravity},      
Int. J. Mod. Phys. {\bf A8} (1993) 1125.}
\lref\WestHowe{P. Howe and P. West, \nihil{The Complete 
N=2 D=10 Supergravity}, Nucl. Phys. {\bf B238} (1984) 181.}
\lref\BalKr{V. Balasubramanian, P. Kraus and A. Lawrence,
\nihil{Bulk vs. boundary dynamics in anti-de Sitter spacetime},
Phys. Rev. {\bf D59} (1999) 046003, \eprt{hep-th/9805171}.}
\lref\ArgDou{P. C. Argyres and M. R. Douglas,
\nihil{New phenomena in SU(3) supersymmetric gauge theory},
Nucl. Phys. {\bf B448} (1995) 93, \eprt{hep-th/9505062}.} 
\lref\ArgWit{P.~C.~Argyres, M.~Ronen Plesser, N.~Seiberg and
E.~Witten,
\nihil{New N=2 Superconformal Field Theories in Four Dimensions,}
Nucl.\ Phys.\  {\bf B461} (1996) 71, \eprt{hep-th/9511154}.}
\lref\BehCv{K.~Behrndt and M.~Cvetic,
\nihil{Supersymmetric domain wall world from D = 5 simple gauged
supergravity},
\eprt{hep-th/9909058}.}
\lref\Beh{K.~Behrndt,
\nihil{Domain walls of D = 5 supergravity and fixpoints of N = 1 super
Yang-Mills}, \eprt{hep-th/9907070}.}
\lref\SeibWit{N. Seiberg and E. Witten, 
\nihil{Monopole Condensation, and Confinement in N=2 
Supersymmetric Yang-Mills Theory}, Nucl. Phys. {\bf B426} (1994) 19,
\eprt{hep-th/9407087};
\nihil{Monopoles, Duality and Chiral Symmetry Breaking in 
N=2 Supersymmetric QCD}, Nucl. Phys. {\bf B431} (1994) 484,
\eprt{hep-th/9408099}.}
\lref\PetZaf{M.~Petrini and A.~Zaffaroni,
\nihil{The holographic RG flow to conformal and non-conformal theory},
\eprt{hep-th/0002172}.}
%
%%%%%%%%%%%%%%%%%%%%%%%%%%%%%%%%%%%%%%%%%%%%%%%%%%%%%%%%%%%%
  
\Title{
\vbox{
\hbox{CITUSC/00-18}
\hbox{USC-00/02}
\hbox{\tt hep-th/0004063}
}}
{\vbox{\vskip -1.0cm
\centerline{\hbox{N=2 Supersymmetric RG Flows and the IIB Dilaton}}
\vskip 8 pt
\centerline{\hbox{ }}}}
\vskip -.3cm
\centerline{Krzysztof Pilch and Nicholas P.\ Warner }
\medskip
\centerline{{\it Department of Physics and Astronomy}}
\centerline{{\it and}}
\centerline{{\it CIT-USC Center for Theoretical Physics}}
\centerline{{\it University of Southern California}}
\centerline{{\it Los Angeles, CA 90089-0484, USA}}

\bigskip
\bigskip
We show that there is a non-trivial relationship between the dilaton
of IIB supergravity, and the coset of scalar fields in
five-dimensional, gauged $\cN=8$ supergravity.  This has important
consequences for the running of the gauge coupling in massive
perturbations of the AdS/CFT correspondence.  We conjecture an exact
analytic expression for the ten-dimensional dilaton in terms of
five-dimensional quantities, and we test this conjecture.
Specifically, we construct a family of solutions to IIB supergravity
that preserve half of the supersymmetries, and are lifts of
supersymmetric flows in five-dimensional, gauged $\cN=8$ supergravity.
Via the AdS/CFT correspondence these flows correspond to softly broken
$\cN=4$, large N Yang-Mills theory on part of the Coulomb branch of
$\cN=2$ supersymmetric Yang-Mills.  Our solutions involve non-trivial
backgrounds for all the tensor gauge fields as well as for the dilaton
and axion.

\vskip .3in
\Date{\sl {April, 2000}}
%\draft

%
\parskip=4pt plus 15pt minus 1pt
\baselineskip=15pt plus 2pt minus 1pt

\newsec{Introduction}

It has been evident over the last year that five-dimensional
supergravity theories are very powerful tools in the study of the
AdS/CFT correspondence \refs{\JMalda,\SGIKAP,\WitHolOne}.  In
particular,
gauged $\cN=8$ supergravity in five dimensions \refs{\GRW,\PPvN}
describes $\cN=4$ Yang-Mills theory in the large $N$ limit under
perturbations that involve fermion or scalar bilinear operators
\refs{\SGIKAP,\WitHolOne,\DistZam,\GPPZold,\KPW}.  
What has been less evident is exactly how the five-dimensional
solutions are lifted to ten-dimensional solutions.  An example of such
a lift was given in \PWcrpt\ for the non-trivial, supersymmetric
critical point of \KPW, however, as yet, no non-trivial lifts of
massive five-dimensional flows have been obtained.  One of the
purposes of this paper is to give the exact ten-dimensional solution
for the five-dimensional $\cN=2$ supersymmetric flows.

The other, and more far-reaching purpose of this paper is to solve a
beautiful subtlety in consistent truncation: a subtlety that has
significant consequences for the field theory side of the AdS/CFT
correspondence.  Specifically, there is an $SL(2,\IR)$ invariance of
the five-dimensional gauged supergravity theory, and perturbatively,
the coset, $SL(2,\IR)/SO(2)$ corresponds to the
ten-dimensional dilaton and axion.  Combining these facts, it is a
natural conflation to assume that this is always true: \ie\ that the
$SL(2,\IR)$ invariance in the five-dimensional theory ``sweeps out''
the ten-dimensional dilaton/axion coset.  This turns out to be false,
and indeed false in a very interesting way.

The scalars of $\cN=8$ supergravity are decribed by the coset
$E_{6(6)}/USp(8)$.  In terms of $SO(6)$ representations, the
non-compact generators constitute the ${\bf 20'} \oplus {\bf 10}
\oplus \overline{\bf 10}  \oplus {\bf 1 }  \oplus {\bf 1 }$.
The two singlets are dual to the gauge coupling and theta-angle, while
the ${\bf 20'}\oplus {\bf 10}
\oplus \overline{\bf 10}$ are respectively dual to the Yang-Mills
scalar and fermion bilinears:
\eqn\dualops{{\rm Tr}\,(X^A X^B)  \,-\, {1 \over 6} \,\delta^{AB}\,
{\rm Tr}\,(X^C X^C) \, , \qquad
{\rm Tr}\,(\lambda^i \lambda^j)\ ,  \qquad
{\rm Tr}\,(\bar \lambda^{\bar i} \bar \lambda^{\bar j}) \ .}
The subgroup $SL(6,\IR) \times SL(2,\IR) \subset E_{6(6)}$ describes
the Yang-Mills theory on the Coulomb branch \refs{\BalKr,\FGPWb}, or
under purely scalar mass perturbations.  In this sector the
$SL(2,\IR)$ factor can indeed be identified with the ten-dimensional
dilaton/axion coset.  However, for more general $E_{6(6)}$ matrices,
\ie\ when fermion masses or vevs are non-zero it turns out that the
relationship is far from trivial.  Indeed, in this paper we conjecture
that the $SL(2,\IR)$ matrix, $S$, that parametrizes the
ten-dimensional IIB dilaton/axion is related to the scalar $E_{6(6)}$
matrix, $\cV$, of five-dimensional supergravity via:
\eqn\SAnsprelim{\Delta^{-{4 \over3}}\, (S\,S^T)^{\alpha \beta} \,=\,
{\rm const}\,\times\,\epsilon^{\alpha \gamma} \epsilon^{\beta \delta}\,
\cV_{I \gamma}{}^{ a b} \,\cV_{J \delta}{}^{ c d}
\,x^I x^J\, \Omega_{ac}\, \Omega_{bd} \ .}
In this equation $x^I$ are the cartesian coordinates on the
compactification $5$-sphere: $\sum_I (x^I)^2 =1$, and $\Delta $ is
related to the determinant of the internal metric.  One may also
define $\Delta $ by taking the determinant of both sides of
\SAnsprelim\ and using the unimodularity of $S$.

The five-dimensional scalar potential is invariant under $SL(2,\IR)$,
and this is broken to $SL(2,\ZZ)$ in the string theory.  The
non-compact generators of this $SL(2,\IR)$ are thus naturally
identified with the $\cN=4$ gauge coupling.  Equation \SAnsprelim\
thus shows that the gauge coupling and theta-angle on the branes is a
non-trivial, but exactly known function of the fermion and scalar
masses and vevs, and of the $\cN=4$ gauge coupling.  Indeed, most of
the five-dimensional flows considered to date have constant values for
the five-dimensional ``dilaton and axion.''  Equation \SAnsprelim\
gives precisely the non-trivial running of the coupling for such flows,
and presumably for $N=1$ supersymmetric flows it should subsume an
integrated version of the NSVZ exact beta-function \NSVZ.

We will examine some of these ideas in this paper, and test
the conjecture \SAnsprelim\ by considering the $\cN=2$ supersymmetric
flow.  In this flow, $\cN=4$ supersymmetric Yang-Mills is softly broken
to $\cN=2$ by introducing a mass for the adjoint hypermultiplet.
Gauged $\cN=8$ supergravity also allows us to probe a lowest mode of
the Coulomb branch of the $\cN=2$ theory: there is a supergravity
scalar that corresponds to turning Yang-Mills scalar vevs that, in the
absence of the fermion mass, corresponds to spreading out the branes
into a uniform disk distribution.  For this flow, \SAnsprelim, in
principle, gives a supergravity prediction for
$\tilde \tau(\tau, m, u)$, where $\tilde \tau$ is the running $\cN=2$
coupling, $\tau$ is the $\cN=4$ coupling, $m$ is the fermion mass and
$u$ is the non-trivial scalar vev.  As we will see there
are some subtleties yet to be understood.

We begin in section 2 by reviewing the five-dimensional
description of the $\cN=2$ supersymmetric flow, and we compute
the running of the dilaton predicted by  \SAnsprelim. In section
3 we use the results of \KPW\ to give the exact ten-dimensional
metric for the flow, and then we examine the linearized Ansatz
for the ten-dimensional $2$-form fields.  We then show that
the $\cN=2$ flow must necessarily involve a running ten-dimensional
dilaton.  In section 4 we obtain the complete ten-dimensional
solution, confirming our prediction of the dilaton/axion behaviour.
Finally, in an appendix we give the consistent truncation argument
that led us to the formula \SAnsprelim.

\newsec{The $N=2$ RG flow in five dimensions}

The flow that preserves $\cN=2$ supersymmetry can be obtained from the
superpotentials considered in \FGPWa.  We need to turn on the
supergravity scalar fields dual to the operators:
\eqn\twoflowops{ \cO_b \eql \sum_{j=1}^4\,  {\rm Tr} \big( X^j
X^j \big)  \,-\, 2\sum_{j=5}^6 {\rm Tr}\big( X^{j} X^{j})   \,, \qquad
\cO_f \eql    {\rm Tr} \big( \lambda^3 \lambda^3 + \lambda^4
\lambda^4 \big)  \,,}
and to the complex conjugate operator, $\overline \cO_f$.
In the conventions of \FGPWa, the corresponding supergravity scalars
are $\alpha$ and $\chi = \varphi_1 =\varphi_2$.  On this
subector the tensor $W_{ab}$ has two distinct eigenvalues,
each with degeneracy $4$.  One of these two eigenvalues provides a
superpotential for the flow:\foot{Note that the eigenvalue that provides
the superpotential for the flow considered here is not the same
as the eigenvalue that provided the $\cN=1$ flow in \FGPWa.}
\eqn\superpot{W\eql  - {1 \over \rho^2}\,-\,
{1 \over 2}\,\rho^4\, \cosh(2 \chi) \ ,}
where, as usual, $\rho = e^\alpha$, and where the potential is given by:

\eqn\Ntwopot{\eqalign{\cP \eql & -{g^2 \over 4}\, \rho^{-4}\,-\, {g^2
\over 2}\, \rho^2  \, \cosh(2 \chi)  \,+\, {g^2   \over 16}\,\rho^8\,
\sinh^2(2 \chi) \cr
\eql & \  {g^2\over 48} \Big({\del W \over \del \alpha} \Big)^2 \,+\,
{g^2\over 16} \Big({\del W \over \del \chi} \Big)^2\,-\,
{g^2\over 3}\,W^2  \,. }}
The kinetic term on this sector is: $ -3(\del \alpha)^2 - (\del
\chi)^2$.

Taking the flow metric to have the form
\eqn\RGFmetric{
ds^2_{1,4} = e^{2 A(r)} \eta_{\mu\nu} dx^\mu dx^\nu - dr^2 \,,}
one then finds that supersymmetric flow equations are:
\eqn\floweqs{\eqalign{{d \alpha \over d r} \eql  &{g \over 12}\,
{\del W \over \del \alpha} \eql {g \over 6}\,\Big({1 \over \rho^2}\,
-\, \rho^4\, \cosh(2 \chi) \Big) \ , \cr
{d \chi \over d r} \eql  &{g \over 4}\, {\del W \over
\del \chi} \eql -{g \over 4}\,  \rho^4\, \sinh(2 \chi)  \ ,}}
along with the auxilliary equation
\eqn\cosmic{{d A \over d r}\eql  - {g \over 3}\,W \ .}

The first step to solving the system \floweqs,\cosmic\ is to write
everything as a function of $\chi$.  Thus:
\eqn\alphachi{ {d \alpha \over d \chi} \eql  -{2 \over 3}\, \bigg(
{1 \over \rho^6 \sinh(2 \chi)}\,-\, {\cosh(2 \chi) \over
\sinh(2 \chi)} \bigg) \ ,}
and
\eqn\Achi{ {d A \over d \chi} \eql  -{2 \over 3}\, \bigg(
{2 \over \rho^6 \sinh(2 \chi)}\,+\, {\cosh(2 \chi) \over
\sinh(2 \chi)} \bigg) \ .}
First note that:
\eqn\Achi{ {d  \over d \chi}\,(A-2\alpha ) 
\eql -2\,{\cosh(2 \chi) \over
\sinh(2 \chi)}   \ . }
This is trivially integrated with respect to $\chi$, and it yields
\eqn\Asol{e^A\eql  k {\rho^2 \over  \sinh(2 \chi)} \ ,}
where $k$ is a constant.

To integrate the equation for $\alpha$, simply define
$\beta = \alpha - {1 \over 3} \log(\sinh(2 \chi))$ and observe that
$$
{d \beta \over d \chi} \eql  -{2 \over 3}\,
{1 \over \rho^6 \sinh(2 \chi)} \eql    -{2 \over 3}\,
{e^{-6 \beta} \over \sinh^3 (2 \chi)}\ .
$$
It is also elementary to integrate this, and one then finds:
\eqn\alphasol{\eqalign{\rho^6 \eql & \cosh(2 \chi) \,+\, \sinh^2 (2
\chi)\,
\Big[\,\gamma \,+\, \log\Big({\sinh(\chi) \over \cosh(\chi)} \Big)\,
\Big]  \cr  \eql & c  \,+\, (c^2 -1)\,\Big[\,\gamma \,+\, {1 \over
2}\,\log
\Big({c-1 \over c+1} \Big)\, \Big]  \ , }}
where $c = \cosh(2 \chi)$ and $\gamma$ is a constant of integration.

This formula has different asymptotics for $\gamma$ positive, negative
or zero.  Since the superpotential has a manifest symmetry under $\chi
\to -\chi$, we focus on $\chi >0$: If $\gamma$ is positive then
$\alpha \sim {2
\over 3}\chi + {1 \over 6}\log({\gamma \over 4})$ for large (positive)
$\chi$.  If $\gamma$ is negative then $\chi$ limits to a finite value
as $\alpha$ goes to $-\infty$.  If $\gamma =0$ then we get the
interesting ridge-line flow with $\alpha \sim -{ 1 \over 3}\chi + {1
\over 6}\log({4 \over 3})$ for large (positive) $\chi$.  Some of these
flows are shown in Figure 1.

We claim that the choice, $\gamma =0$ corresponds to the pure $\cN=2$
flow with vanishing scalar vev.    The simplest argument
for this claim is obtained from Figure 1.  There are several obvious
ridges in this figure.  The ridges with $\chi =0$ and $\alpha$ varying
are two of the $\cN=4$ supersymmetric Coulomb branch flows identified
in \FGPWb.   The two other ridges are equivalent under $\chi \to -\chi$
and are obtained by setting $\gamma = 0$.  They correspond to massive
supersymmetric flows, and there is only one such ``preferred flow''
namely the  pure $\cN=2$ flow with no scalar vev.
One should note that the flow along the $\alpha$-axis
to the left corresponds to the Coulomb branch in which the branes
spread out in a disk in the $(X^5,X^6)$ directions, whereas the
other direction corresponds to a brane distribution in the
$(X^1,X^2,X^3,X^4)$ directions.  The moduli space that we seek makes
the scalars $(X^1,X^2,X^3,X^4)$ massive, but leaves us with the vevs
in the $(X^5,X^6)$ directions.  Hence it is natural to identify
the softly broken $\cN=4$ theory with the $\gamma=0$
ridge and everything to the left of it.  This is essentially the
argument given in \SGsing, where is was further argued that the flows
to the right of the $\gamma = 0$ should be viewed as unphysical.

%%%%%%%%%%%%%%%%%%%%%%%%%%%%%%%
\goodbreak\midinsert
\vskip .5cm
\centerline{ {\epsfxsize 3in\epsfbox{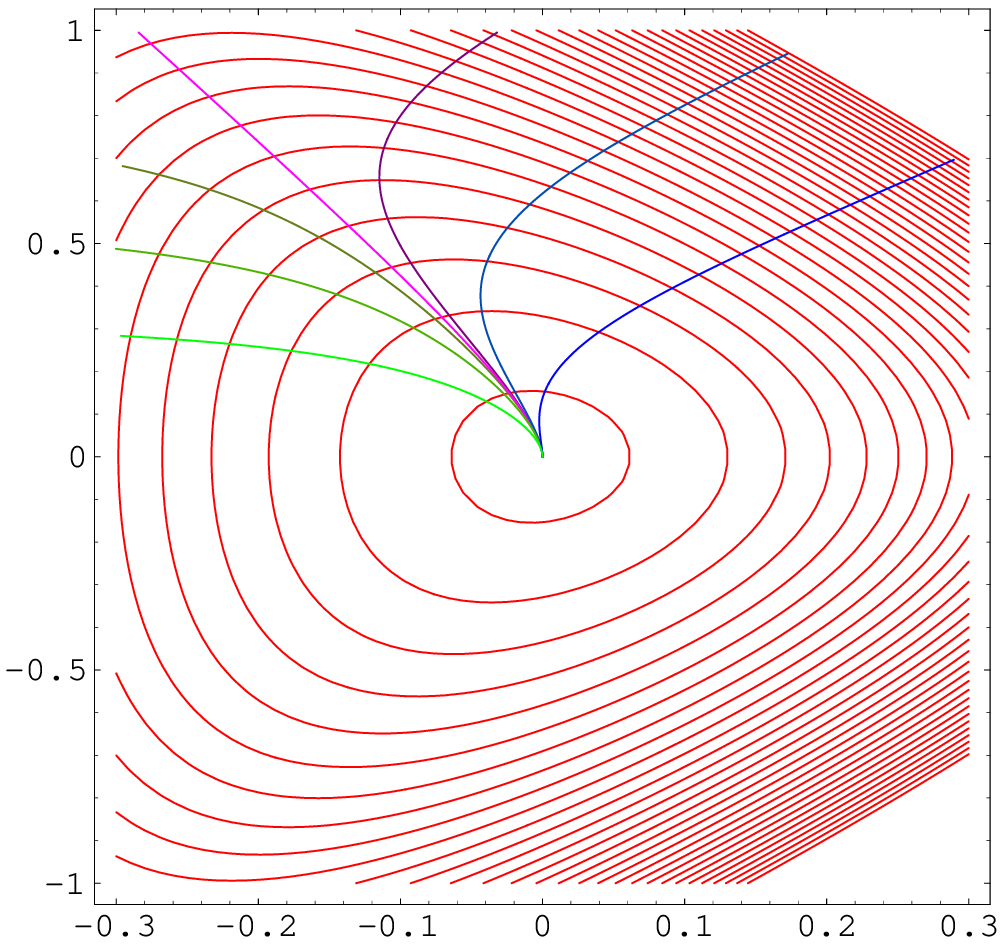}}}
\leftskip 2pc
\rightskip 2pc\noindent{\ninepoint\sl \baselineskip=8pt
{\bf Fig.~1}: Contours of the superpotential showing some of the
steepest descent flows. The horizontal and vertical axis are $\alpha$
and $\chi$, respectively. The middle, ridge-line flow has $\gamma=0$
and the three flows to the left and right have $\gamma<0$ and
$\gamma>0$, respectively. }
\endinsert
%%%%%%%%%%%%%%%%%%%%%%%%%%%%%%%

\newsec{The ten-dimensionsional solution }

In this section we ``lift'' the ${\cal N}=2$ flows to solutions of the
chiral IIB supergravity in ten dimensions \refs{\JSIIB,\WestHowe}.  As
we have already explained above, one should expect to find that all
bosonic fields in ten dimensions are non-vanishing and as a result all
bosonic field equations become nontrivial. Those equations consist of
\JSIIB:
\smallskip
\noindent
$\bullet$\quad  The Einstein equations:
\eqn\tenein{
R_{MN}=T^{(1)}_{MN}+T^{(3)}_{MN}+T^{(5)}_{MN}\,, } where the energy
momentum tensors of the dilaton/axion field, $B$, the three index
antisymmetric tensor field, $F_{(3)}$, and the self-dual five-index
tensor field, $F_{(5)}$, are given by
\eqn\enmomP{
T^{(1)}_{MN}\eql P_MP_N{}^*+P_NP_M{}^*\,,
}
\eqn\enmomG{
T^{(3)}_{MN}\eql
       {1\over 8}(G^{PQ}{}_MG^*_{PQN}+G^{*PQ}{}_MG_{PQN}-
        {1\over 6}g_{MN} G^{PQR}G^*_{PQR})\,,
}
\eqn\enmomF{
T^{(5)}_{MN}\eql {1\over 6} F^{PQRS}{}_MF_{PQRSN}\,.
}
\smallskip

\noindent
We work here in the unitary gauge in which $B$ is a complex scalar
field and
\eqn\defofPQ{
P_M\eql f^2\partial_M B\,,\qquad Q_M\eql f^2\,{\rm Im}\,(
B\partial_MB^*)\,,
}
with
\eqn\defoff{
f\eql {1\over (1-BB^*)^{1/2}}\,,
}
while the antisymmetric tensor field $G_{(3)}$ is given by
\eqn\defofG{
G_{(3)}\eql f(F_{(3)}-BF_{(3)}^*)\,.
}
\noindent
$\bullet$\quad The Maxwell equations:
\eqn\tenmaxwell{
(\nabla^P-i Q^M) G_{MNP}\eql P^M G^*_{MNP}-{2\over 3}\,i\,F_{MNPQR}
G^{PQR}\,.
}
\eqnn\tengsq
$\bullet$\quad The dilaton equation:\foot{A perceptive reader might
have noticed that the sign on the right hand side of our dilaton
equation is opposite to that in (4.11) and (5.1) of \JSIIB. It appears
that there is an error in \JSIIB\ in passing from (4.10) to (4.11). An
independent verification of the sign is to check whether the Bianchi
identity $\nabla^M R_{MN}-{1\over 2}\nabla_N R=0$ is consistent with
the field equations. Indeed, this is the case for the sign in \tengsq,
but not for the one in \JSIIB. The consistency between 
the Bianchi identities and the field equations was 
also verified in \CastPes, with the resulting 
correct sign in the dilaton equation. }
$$
(\nabla^M -2 i Q^M) P_M\eql -{1\over 24} G^{PQR}G_{PQR}\,. \eqno\tengsq
$$
$\bullet$\quad The self-dual equation:
\eqn\tenself{
F_{(5)}\eql *F_{(5)}\,,
}
In addition, $F_{(3)}$ and $F_{(5)}$ satisfy Bianchi identities which
follow from the definition of those field strengths in terms of their
potentials:
\eqn\defpotth{\eqalign{
F_{(3)}&\eql dA_{(2)}\,,\cr
F_{(5)}&\eql dA_{(4)}-{1\over 8}\,{\rm Im}( A_{(2)}\wedge
F_{(3)}^*)\,.\cr}
}

Our strategy for constructing the ten-dimensional solution is to start
with the consistent truncation Ansatz for the metric \KPW. By
examining the resulting Ricci tensor, we arrive at identities that,
together with the $SU(2)\times U(1)^2$ symmetry, essentially determine
the general structure of the antisymmetric tensor fields. The next
crucial step is to solve the linearized Maxwell equation for the three
index tensor field, $G_{(3)}$, which turns out to yield a
non-vanishing source for the dilaton/axion field in \tengsq. Using the
consistent truncation Ansatz for the dilaton/axion we are then able to
completely solve the Einstein equations to all orders, and fine tune
all the phases and constants using the remaining equations of motion
and the Bianchi identities.

\subsec{The metric}

Along the flow the ten-dimensional space-time is topologically a
product of $AdS_5$ and a sphere, $S_5$, with the ``warped product''
metric of the form:
\eqn\warpmetr{
ds_{10}^2\eql \Omega^2 ds_{1,4}^2+ds_5^2\,.
}
where $ds_{1,4}^2$ has been given in \RGFmetric.  The ``internal''
metric, $ds_5^2$, and the warp factor, $\Omega^2$, are determined by
the consistent truncation in terms of the scalar fields as discussed
in the appendix.

The calculation of the explicit form of the internal metric using (A.8)
is essentially the same as in \PWcrpt\ (see also \CLP). We represent
$S_5$ as a unit sphere in $\IR^6$ with the cartesian coordinates
$x^I$, $I=1,\ldots,6$, and pass to suitable spherical coordinates to
make the $SU(2)\times U(1)^2$ symmetry manifest. This is accomplished
by setting
$$ u^1\eql x^1+ix^4\,,\qquad u^2\eql x^2+i x^3\,,\qquad u^3\eql
x^5-ix^6\,, $$
so that $(u^1,u^2)$ transform as a doublet of $SU(2)$ with zero
charge, and $u^3$ is a singlet with charge 1. The remaining $U(1)$
rotates between the doublet and its conjugate.  Then we use the group
action to reparamerize these coordinates as follows:
\eqn\angcoords{
\left(\matrix{u^1\cr u^2\cr}\right)\eql
\cos\theta\, g(\al_1,\al_2,\al_3)\,
\left(\matrix{1\cr 0\cr}\right)\,,\qquad
u^3\eql e^{i\phi}\,\sin\theta\,,}
where $g(\al_1,\al_2,\al_3)$ is an $SU(2)$ matrix expressed in terms of
your Euler angles.

Using explicit scalar 27-beins, $\widetilde {\cal V}_{IJ}{}^{ab}$, for
the flow, and parametrizing the Killing vectors in terms of the
coordinates above we arrive at the final result for the internal metric:

\eqn\intmetr{\eqalign{
ds_5^2 & \eql
{a^2\over 2} { (cX_1X_2)^{1/4}\over\rho^3} \left(
 c^{-1} d \theta^2 +\rho^6\cos^2\theta\,\Big({\sigma_1^2\over cX_2}
+{\sigma_2^2+\sigma_3^2\over X_1}\Big)+\sin^2\theta\, 
{d\phi^2\over X_2} \right)
\cr } }
where, as in section 2, $ c \equiv \cosh(2 \chi )$ and
$\rho  \equiv e^{ \alpha }$.
The warp factor is given by:
\eqn\warpfac{
\Omega^2\eql \Delta^{-{2\over 3}}\eql {(cX_1X_2)^{1/4}\over \rho}\,. }
The two functions, $X_1$ and $X_2$, are defined by
\eqn\thexs{\eqalign{
X_1(r,\theta)&\eql \cos^2\theta + \rho(r)^6 \cosh(2\chi(r))
\sin^2\theta\,,\qquad\cr
X_2(r,\theta)&\eql \cosh(2\chi(r))\cos^2\theta+
\rho(r)^6\sin^2\theta\,.\cr}
}
and we have introduced the constant, $a$, to account for the arbitrary
normalization of the Killing vectors. As usual, $\sigma_i$, $i=1,2,3$,
are the $SU(2)$ left-invariant forms, satisfying $d\sigma_i=2
\sigma_j\wedge \sigma_k$.

Clearly, the metric \intmetr\ is invariant under $SU(2)$ and the two
$U(1)$'s, where the first one, $U_\phi(1)$, acts by a translation in
$\phi$, while the second second one, $U_{23}(1)$, rotates $\sigma_2$
into $\sigma_3$.

Note that in our coordinates the metric \warpmetr\ is almost diagonal.
We choose the corresponding orthonormal frames $e^M$, $M=1,\ldots,10$,
\eqn\vielbeins{\eqalign{e^1&\propto dx^0\,,\qquad e^2 \propto
dx^1\,,\qquad e^3 \propto dx^2\,,\qquad
e^4 \propto dx^3\,, \qquad e^5 \propto dr\,,\cr
e^6&\propto d\theta\,,\qquad e^7\propto \sigma_1\,,\qquad e^8\propto
\sigma_2\,,\qquad e^9\propto \sigma_3\,,\qquad e^{10}\propto d\phi
\ .  \cr}}

The computation of the Ricci tensor becomes rather involved and is
most conveniently carried out on a computer. We find that the only
non-vanishing off-diagonal components are $R_{56}=R_{65}$, while the
diagonal components satisfy  the obvious identities, $R_{11}=
-R_{22}= -R_{33}= -R_{44}$ and $R_{88}=R_{99}$, which follow from the
symmetries of the metric.  We also find the rather non-trivial
identity:
\eqn\ricciid{
R_{77}+R_{88}\eql 2 R_{11}\,.}
We will see below that this equation implies the vanishing
of some components of the $3$-form field strengths.

Throughout the calculation we use the flow equations \floweqs\ to
eliminate
derivatives with respect to $r$, so that the final result depends on
rational functions of the scalar fields $c(r)$ and $\rho(r)$ and
trigonometric functions of $\theta$. While most of the components of
the Ricci tensor are sufficiently complicated to prevent us from
reproducing them here, we note that the combination
$R_{10\,10}-R_{77}$ is rather simple. This will turn out important for
solving the Einstein equations below.

\subsec{The dilaton}

We now use \SAnsprelim\  to obtain the  $SL(2,\IR)
\equiv SU(1,1)$ scalar matrix of the ten-dimensional type IIB
theory.  We find that on the $(\alpha,\chi)$ parameter
space considered in section 2,  the right-hand side of
\SAnsprelim\ yields:
\eqn\Vielmat{ {1 \over \rho^2}\, \left(\matrix{ c X_1 \cos^2 \phi +
X_2 \sin^2 \phi & \rho^6 s \sin^2 \theta \, \sin\phi \cos\phi
  \cr   \rho^6 s \sin^2 \theta \,\sin\phi\cos\phi &
c X_1 \sin^2\phi + X_2 \cos^2\phi } \right) \, ,}
where $c = \cosh(2 \chi)$ and $s = \sinh(2 \chi)$.
According to  \SAnsprelim\ the determinant should yield
$\Delta^{-8/3}$, and from this we obtain:
\eqn\DeltaNtwo{ \Delta \eql   {\rho^{3/2} \over \big(c\,X_1\,X_2
\big)^{-3/8}} \, ,}
which is consistent with \warpfac.  Note that $\Delta^2 \equiv {\rm
det}\,( g_{mp}\,{\displaystyle \gop}{}^{pq})$ where $g_{mp}$ is the
internal metric given by \intmetr\ and ${\displaystyle \gop}{}^{pq}$
is the inverse of the ``round'' internal metric at $\chi=\alpha=0$.
One can determine $S$ in the symmetric gauge by taking the square root
of this matrix.  For direct comparison with the IIB field equations in
\JSIIB\ we also pass to the $SU(1,1)$ basis in which the dilaton/axion
matrix takes the form:
\eqn\Vmat{ V \eql  f\, \left(\matrix{ 1 & B \cr B^* &1 } \right) \, .}
We then find the following result:
\eqn\fBdefns{ f \eql  \half\,\Big(  \Big( {c X_1 \over X_2 }
\Big)^{1\over 4} + \Big( {c X_1 \over  X_2 }\Big)^{-{1\over 4}}
\Big)\, , \qquad f B \eql  \half\,\Big(  \Big( {c X_1 \over X_2 }
\Big)^{1\over 4} - \Big( {c X_1 \over  X_2 }\Big)^{-{1\over 4}}
\Big)\,e^{2 i \phi} \, .  }

\subsec{The antisymmetric tensor fields}

A Poincare and $SU(2)\times U(1)^2$ invariant Ansatz for the self-dual
antisymmetric tensor field, $F_{(5)}$, reads
\eqn\theFten{
F_{(5)}\eql {\cal F}+*{\cal F}\,, \qquad
{\cal F}\eql dx^0\wedge dx^1\wedge dx^2\wedge dx^3\wedge dw}
where $w(r,\theta)$ is an arbitrary function. The self-duality
equation \tenself\ is then satisfied by construction.  The
non-vanishing components of the energy momentum tensor satisfy:
\eqn\emFFten{
T_{11}^{(5)}\eql -T_{22}^{(5)}\eql\ldots\eql
-T_{33}^{(5)}\eql T_{77}^{(5)}\eql\ldots\eql
T_{10\,10}^{(5)} \eql {\cal A}^2+{\cal B}^2\,,
}
\eqn\emFFtwo{
T_{55}^{(5)} \eql - T^{(5)}_{66}\eql {\cal A}^2-{\cal B}^2\,,
}
and
\eqn\emFFthr{
T_{56}^{(5)}\eql T_{65}^{(5)}\eql 2 {\cal A}{\cal B}\,,
}
where
\eqn\theab{\eqalign{
{\cal A}&\eql {2^{3/2}\over a k^4}\,
        {s^4 \over c^{1/8}\rho^{9/2}(X_1X_2)^{5/8}}\,
        {\partial w\over\partial\theta}\,,\cr
{\cal B}&\eql { 2\over k^4}\,
        {s^4 \over \rho^{11/2}(c X_1X_2)^{5/8}}\,
        {\partial w\over\partial r}\,,\cr} }
and $k$ is the constant introduced in \Asol.

The most general Ansatz for the potential, $A_{(2)}$, that gives an
$SU(2)\times U_{23}(1)$ invariant field strength, $G_{(3)}$, with the
$U_\phi(1)$ charge $+1$ is
\eqn\Aanz{
A_{(2)}\eql e^{i\phi}\, \big( a_1(r,\theta)\,d\theta\wedge\sigma_1 +
a_2(r,\theta)\,\sigma_2\wedge\sigma_3 + a_3(r,\theta) \,\sigma_1\wedge
d\phi + a_4(r,\theta)\,d\theta\wedge d\phi
\big)\,,}
where $a_i(r,\theta)$ are some arbitrary complex functions.

In solving the equations of motion we want to test the conjecture
\fBdefns, and assume as little as possible of its form.  However,
the group theory implies that the dilaton must have the form:
\eqn\dilform{B\eql  b(r,\theta)\,e^{2 i \phi} \ .}
In particular, the energy-momentum tensor of the dilaton vanishes in
the directions $7,8,9$ as well as in the directions parallel to brane.
It follows that the only energy-momentum tensor that can contribute to
\ricciid\ is $T^{(3)}$.  One finds that the only way to satisfy
\tenein\ and \ricciid\ is to impose:
\eqn\killa{ a_4(r,\theta)\eql 0\,. }

If one computes the components of the energy momentum tensors,
$T^{(1)}_{MN}$ and $T^{(3)}_{MN}$, in the orthonormal frame
\vielbeins\ we find that they have some non-zero off-diagonal
components where the corresponding components of the Ricci tensor
vanish.  These algebraic constraints are most simply (but not
necessarily uniquely) solved by requiring that the phase of $b$ (if
any) is independent of $r$ and $\theta$, and the functions $a_1$ and
$a_2$ are pure imaginary, while $a_3$ is real.

At this point it is instructive to solve the linearized equations of
motion in the UV limit, $r\rightarrow \infty$. In this limit the
metric \warpmetr\ approaches the product metric on $AdS_5\times S_5$
with the radii $L=2/ g$ and $a/\sqrt{2}$, respectively, as seen
from the expansions $A(r)\sim r/L$, $\chi \sim e^{-r/L}$, $\alpha \sim
r e^{-2r/L}$ that follow from \floweqs\ and \cosmic. The five-index
tensor should reduce to the usual Freund-Rubin Ansatz.  Thus we take
\eqn\linearf{
w\eql m{k^4\over 4 s^4} + O({1\over s^2})}
and find that the UV limit reproduces the correct solution of the
Einstein equations at the ${\cal N}=8$ critical point provided
\eqn\aconst{
a^2\eql {8\over g^2}\qquad {\rm and}\qquad m^2\eql {1\over 16}\,.}

Futhermore we consider a linearized Ansatz for the two-index
gauge potentials of the form:
\eqn\linofG{
a_i(r,\theta)\eql e^{-\mu r/L} \,\tilde a_i(\theta) 
+O(e^{-(\mu+1) r/L})
\,, }
for some constant $\mu$.  From the linearized analysis of
\refs{\SGIKAP,\WitHolOne}, and because these linearized fields are
dual to fermion bilinears on the brane, we must have either $\mu=1$ or
$\mu=3$.  The former mode is non-normalizable and corresponds to a
massive flow, while the latter is normalizable and corresponds to a
``gaugino condensate,'' which is a vacuum modulus.

The Maxwell equations do indeed imply that $\mu=1$ or $\mu=3$.
There are also four independent Maxwell equations for the functions
$\tilde a_i$, which can be reduced to a single third order equation
(which has regular singular points) for $\tilde a_1(\theta)$.
The regular solution is
\eqn\regsola{
\tilde a_1(\theta)\eql a_0 \,i\, \cos(\theta)\,, }
where $a_0$ is a real constant. The remaining two functions are
\eqn\otheras{
\tilde a_2(\theta) \eql a_0 \,i\, \cos^2\theta\,\sin\theta\,
\qquad {\rm and }\qquad
\tilde a_3(\theta)\eql -a_0\cos^2\theta\, \sin\theta\,. }
Substituting this solution into the right hand side of the dilaton
equation
we find
\eqn\probwithG{
{1\over 24}G_{MNP}G^{MNP}\,\propto\, (\mu^2 -9)\, e^{-2\mu r/L} \,
e^{2i\phi}
\sin^2\theta\,. }
This reveals two interesting features:  If  $\mu =3$ then the
dilaton does {\it not} flow at lowest order, whereas if $\mu =1$
then the  dilaton  {\it must} flow.  So deforming the ground state
of the Yang-Mills theory does not (at lowest order) require the
dilaton to run, but if the flow involves giving a mass
to the fermions then the dilaton must run.  A similar
linearized analysis has recently been done for some $\cN=1$
supersymmetric flows, and once again it was shown that the
dilaton had to run \PetZaf.

\newsec{The complete solution in ten dimensions}

With the general structure of the solution determined by
group theory and the linearized form, we now turn to the full solution.
We take the metric in
\warpmetr, \RGFmetric\ and \intmetr, the dilaton in \Vmat\ and \fBdefns,
the five-index tensor field, $F_{(5)}$, of the form \theFten\ and the
three-index tensor, $G_{(3)}$, defined by the potential $A_{(2)}$ in
\Aanz\ with pure imaginary functions  $a_1$ and $a_2$,  a real
function $a_3$ and $a_4=0$.

\subsec{Solving the equations of motion}

We start with the Einstein equations \tenein\ and cosider linear
combinations for which there are some cancellations or obvious
simplifications of the energy momentum tensors on the right hand side.

The first such example is the difference between the $(10,10)$ and
$(1,1)$ equations, where we find
\eqn\nonzeroric{
R_{10\,10}-R_{11}\not=0\,.}
Now, it is easy to see that the reality conditions on the functions
$a_i$ imply the identity
\eqn\gtenid{
T^{(3)}_{10\,10}-T^{(3)}_{11}\eql {e^{-2i\phi}\over 24}
\, G_{MNP}G^{MNP}\,.  }
Combined with the dilaton equation \tengsq, this provides us with a
nontrivial test of the consistent truncation Ansatz for the
metric and the dilaton, namely
\eqn\firsttest{
R_{10\,10}-R_{11}\eql 2 |P_{10}|^2 +  e^{-2i\phi} (\nabla^M-2 i Q^M)
P_M\,.
}
We find that this identity is indeed satisfied by the dilaton/axion
given in \fBdefns.

In fact, the above calculation can be viewed as an independent
confirmation that the ten-dimensional dilaton must run. The
only imput that goes into \gtenid\ is a symmetry constraint on
which components, $G_{MNP}$, can be non-zero and that those components
are purly real or imaginary up to the $e^{i\phi}$ phase. Thus, if we
tried to set the dilaton to zero, we would end up with the vanishing
right hand side in \firsttest, which would contradict \nonzeroric.

The next combination of the Einstein equations is  the difference
\eqn\difeincr{
R_{10\,10}-R_{77}-2 |P_{10}|^2\eql T^{(3)}_{10\,10}-T^{(3)}_{77}\,,
}
where we have used $T^{(1)}_{77}=0$ and $T^{(5)}_{10\,10}
=T^{(5)}_{77}$. Evaluating \difeincr\ explicitly we get
\eqn\thedifforg{
{g^2\over 4}{\rho^{-3}\,\sinh^2(2\chi)\, (2 X_1+\rho^6 X^2)^2\over
        (\cosh(2\chi)X_1X_2)^{5/4}} -
{g^2\over 4}    {\rho^9\tanh^2(2\chi)\over
(\cosh(2\chi)X_1X_2)^{5/4}}\eql
|G_{8910}|^2-|G_{567}|^2\,,
}
and verify that in the linearized limit the first and the second term
on the left hand side reduce to the first and the second term,
respectively, on the right hand side. Thus we set
\eqn\theggs{\eqalign{
G_{567}& \eql i\, {g\over 2}\, {\rho^{9/2}\tanh(2\chi)\over
        (\cosh(2\chi)X_1X_2)^{1/8}}\,e^{i\phi} \,,\cr
G_{8910}& \eql -{g\over 2}\, {\sinh(2\chi)(2 X_1+\rho^6 X_2)\over
\rho^{3/2}(\cosh(2\chi)X_1X_2)^{5/8}}\, e^{i\phi} \,,\cr
 }}
where the signs have been chosen to agree with the linearized solution
to the Maxwell equations.

By expanding \defofG\ in terms of the potential and using
\fBdefns\ one obtains
\eqn\thegsofa{\eqalign{
G_{567}&\eql i\, {g^2\over 4} e^{i\phi} {\rho^{1/2} c^{3/4} {\rm
sec}\,\theta
\over (cX_1X_2)^{1/8}}\, {\partial a_1\over \partial r}\,,\cr
G_{8910}&\eql -{g^3\over 8}e^{i\phi} {X_1 X_2 \,
{\rm csc}\,\theta \,{\rm sec}^2\theta\over
\rho^{3/2} (cX_1X_2)^{5/8}}\,   (a_2-2 a_3)\,.\cr}
}
The first order equation for $a_1(r,\theta)$ is can be easily
integrated, and by invoking once more the
linearized limit we can identify the remaing two functions
$a_2(r,\theta)$ and $a_3(r,\theta)$. The result is a simple
modification of the linearized solution \regsola\ and \otheras:
\eqn\thesolutiona{\eqalign{
a_1(r,\theta)&\eql -i\,{4\over g^2}\, \tanh(2\chi)\, \cos\theta\,,\cr
a_2(r,\theta)&\eql i\, {4\over g^2}\,  {\rho^6 \sinh(2\chi)
\over X_1}\,
\sin\theta\cos^2\theta\,,\cr
a_3(r,\theta)&\eql {4\over g^2} \,{\sinh(2\chi)\over
X_2}\, \sin\theta\cos^2\theta
\,.\cr}
}

At this stage a nontrivial check for our solution is the sum of the
(5,5) and (6,6) Einstein equations in which according to \emFFtwo\ the
contribution from the yet undetermined energy momentum tensor of the
five-index tensor cancels.

Now, the five-index tensor is calculated using, e.g., the (1,1) and (5,5)
Einstein equations and fixing the sign to agree with the linearized
Ansatz. The result is
\eqn\thefivef{
w(r,\theta)\eql {k^4\over 4} {\rho^6 X_1\over \sinh^2(2\chi)}\,.
}

Finally, it is a matter of a straightforward algebra  to verify
that all the remaining equations of motion and the Bianchi identities
are
satisfied!  The conjecture \SAnsprelim\ has thus passed a collection
of non-trivial tests perfectly.

\subsec{Asymptotic behaviour of the solution}

As was noted in section 2, there are three distinct
asymptotic flows:

\item{(i)} $ \gamma>0\,$:\ $\alpha \sim {2
\over 3}\chi + {1 \over 6}\log({\gamma \over 4})$ for large (positive)
$\chi$.
\item{(ii)} $ \gamma<0\,$:\  $\alpha \to -\infty$,  $\chi \to {\rm const}.$
\item{(iii)} $\gamma =0\,$:\ $\alpha \sim -{ 1 \over 3}\chi + {1
\over 6}\log({4 \over 3})$ for large (positive) $\chi$.

Here we will focus primarily on the last option as we believe it should
exhibit the most interesting new behaviour.  As discussed
earlier, the $ \gamma>0$ flow is expected to be
unphysical, while the  $ \gamma<0$ flow will be akin
to the Coulomb branch of the $\cN=4$ theory.

First, the $ \gamma=0$ is most interesting in that the dilaton
depends upon $\theta$ in the asymptotic limit.  For
$ \gamma<0$ we find $B \to 0$ as $\alpha \to - \infty$, and the
asymptotic dilaton/axion configuration is trivial.
Indeed, it approaches the constant dilaton background of the ${\cal
N}=4$ UV fixed point.
For $ \gamma>0$ we have $B \to  e^{2 i \phi}$
as $\chi \to \infty$ and the matrix $S$ diverges in no matter
what direction we approach the ``core'' of the solution.
For $\gamma=0$ we have:
\eqn\asymB{B~=~ \bigg({{(1 + {2 \over 3} \tan^2(\theta))^{1\over 2}
{}~-~ 1}  \over {(1 + {2 \over 3} \tan^2(\theta))^{1\over 2} ~+~ 1}}
\bigg)~ e^{2i\phi} \,.}
Note that generically $|B| <1$, except for $\theta = \pi/2$
for which one has $B= e^{2 i \phi}$.
So the transition from $\gamma<0$ to   $\gamma>0$ can
be thought of as moving from an asymtotically trivial
dilaton, to a non-trivial dilaton matrix that is finite
except on the ``ring'' $\theta = \pi/2$, and thence to
a dilaton matrix that is asymptotically singular in all
directions.

One should note that $\theta = \pi/2$  corresponds to
setting the cartesian coordinates $x^1=x^2 = x^3 =x^4 =0$
on $S_5$.  The remaining non-trivial coordinates $x^5$
and $x^6$ thus define a ring, which is presumably the
enhan\c{c}on ring of \JPP.  The fact that the dilaton is
asymtotically trivial for $\gamma<0$, is singular for
$\gamma>0$, and exhibits the milder ring singularity
for $\gamma=0$ further supports the identification of
the supergravity flows with the various field theory limits.

The Einstein metric  behaves similarly.  Setting $\gamma=0$
we find that as $\chi \to \infty$ the vielbein behaves
according to:
\eqn\asympviel{\eqalign{ e^a ~\sim~ & 2 \nu~e^{-2 \chi}~dx^a\,,
\quad a =1, \dots,4\,;\qquad e^5 ~\sim~   - 3 a~\nu~ d\chi\,, \qquad
e^6 ~\sim~ \sqrt{ \coeff{3}{2}}~a~\nu~ d\theta \,, \cr
e^7 ~\sim~& \coeff{1}{2\sqrt{2}}~a~ \nu~e^{-2 \chi}~\sigma_1\,,\qquad
e^8 ~\sim~  \coeff{1}{2\sqrt{2}}~a~\nu~
(1 + \coeff{2}{3} \tan^2(\theta))^{-{1\over 2}}~\sigma_2\,, \cr
e^9 ~\sim~ & \coeff{1}{2\sqrt{2}}~ a~\nu~
(1 + \coeff{2}{3} \tan^2(\theta))^{-{1\over 2}}~\sigma_3\,, \qquad
e^{10} ~\sim~  \sqrt{ \coeff{3}{2}}~a~\nu~ \tan(\theta)~d\phi\,,}}
where
\eqn\nudefn{\nu~\equiv~ (\coeff{2}{3})^{1\over 4}~ (\cos
\theta)^{1\over 2}~(1 + \coeff{2}{3} \tan^2(\theta))^{1\over 8} \,.}
As $\chi \to \infty$, and for $\theta \not= \pi/2$,  the metric
remains regular, and the $D3$-branes appear be at the bottom of
an infinitely long throat, much as they are
at a conformal fixed point.  The metric does not quite have
asymptotic conformal invariance owing to the $\chi$-dependence
of $e^7$.  It is however tempting to speculate that the
``near--conformality'' of this asymptotic metric may be related
to a flow to the large $N$ versions of Argyres--Douglas
points \refs{\ArgDou,\ArgWit}.

At $\theta = \pi/2$ the metric (as well as the dilaton) become
singular.  One can re-analyse the asymptotics, and the precise
details depend upon whether one looks at the Einstein metric
or string metric.  The latter still sees an infinitely long
throat, whereas the Einstein metric is singular at finite
distance.  We also find that in either metric, the coefficient
of $d \phi$, and hence the diameter of the
``ring'' singularity goes to {\it infinity} as $\chi \to
\infty$.

\newsec{Conclusions}

In obtaining and checking \SAnsprelim, we have exposed a
very interesting aspect of consistent truncation, whose
physical consequence is that the five-dimensional ``dilaton
coset'' should be identified with the $SL(2,\ZZ)$-invariant
$\cN=4$ coupling, and not with coupling in the gauge theory
on the brane.  Indeed, our expression gives the coupling
on the brane as a function of the  $\cN=4$ coupling and of the
masses and vevs captured by gauged $\cN=8$ supergravity.

Many of the the flows obtained in the literature to date
\refs{\DistZam,\GPPZold,\FGPWa,\FGPWb,\GPPZ,\Beh,\BehCv} 
keep the five-dimensional dilaton fixed.
For these flows \SAnsprelim\ yields the flow of the
dilaton as a function the masses and vevs that drive that
flow, and if the flow is supersymmetric \SAnsprelim\ must
capture the NSVZ exact beta function.   Indeed, it is, in
hindsight rather easy to identify the IIB supergravity
version of the NSVZ exact beta function:  The supersymmetry
variation of the ten-dimensional fermion is \JSIIB:
\eqn\lamvar{\delta \lambda ~=~ \coeff{i}{\kappa}~
\gamma^\mu  P_\mu~\epsilon^* ~-~ \coeff{i}{24} ~G_{\mu \nu \rho}
\gamma^{\mu \nu \rho}~\epsilon\, .}
This must vanish along supersymmetric flows. To linear
order $G_{\mu \nu \rho}$ are the fermion masses, while
$P_r$ is the running coupling.

The foregoing identification is, however, a little superficial in that
it glosses over the fact the dilaton doesn't just depend upon the
radius, $r$.  It also depends upon other coordinates.  For the $\cN=2$
flow considered here it depends upon $\theta$ and $\phi$ (the latter
being very simple).  This makes physical sense in that one starts with
an $\cN=4$ theory, and the flow must ``know'' which directions are
``getting massive.''  To be more specific, Seiberg and Witten
\SeibWit\ showed that there was no infra-red gauge enhancement on the
Coulomb branch of ${\cal N}=2$ theories. Thus, even in the far
infra-red, the brane description of the ${\cal N}=2$ supersymmetric
limit must involve a ``disk-like'' distribution of branes, or at least
something with a corresponding multipole moment. Thus it is entirely
to be expected that the metric and dilaton have non-trivial dependence
on $\theta$ and $\phi$ in the IR limit. This does, however, beg the
question as to how the direction of approach is seen purely from the
perspective of the brane and, in particular, in terms of the
Seiberg-Witten effective action.  More generally, for $\cN=1$ flows,
in which several fields are given independent masses, one would like
to relate the direction on $S_5$ with the physics on the brane and
thereby isolate the running coupling of \NSVZ.

As yet we do not have definitive answers to these issues, but we have
computed some of the dilaton flows for other known supersymmetric
flows.  First, and rather surprisingly, the dilaton and axion are
constant everywhere on the two-parameter $(\alpha,\chi)$ space
underlying the flow of \FGPWa.  On the other hand, the dilaton and
axion {\it do} flow in a very non-trivial way for the flow of \GPPZ.
This is presently under investigation.

Finally, and rather ironically, prior to this work we expended much
effort in trying to find supersymmetric flows in which the
five-dimensional dilaton flows along with other fields. Our failed
efforts, and a heuristic argument based upon energy suggest that there
should be a no-go theorem for such supersymmetric flows.  It would be
interesting to try to prove such a result, and then \SAnsprelim\ would
truly be the unique expression for the running coupling on the brane.
\bigskip

\bigskip
\leftline{\bf Acknowledgements}

We would like to thank  L.~Castellani, S.~Gubser, 
J.~Polchinski and E.~Witten for helpful conversations.  
This work was supported in part by funds
provided by the DOE under grant number DE-FG03-84ER-40168.

\appendix{A}{Consistent truncation revisited}

The central issue in consistent truncation is to determine how the
fields and field equations of the lower-dimensional theory are
embedded in those of a higher-dimensional theory.  One of the keys to
this is to use the gauge invariances of both theories and
supersymmetry transformations to make this mapping precise \MetAns.  In
particular, this is how one can find the exact metric of the
higher-dimensional theory from the metric and scalar fields of the
low-dimensional theory.  

The consistent truncations has been carried out in full detail for the
reduction of the eleven dimensional supergravity to four dimensions
(see, e.g., \refs{\MetAns,\CLP,\Popeetala} and the references therein)
and to seven dimensions \refs{\Trunseven,\NastVam}. The consistent
truncation of IIB supergravity has been analyzed only for various
subsectors of the theory (see, e.g., \refs{\KPW,
\Cetal\CGLPx\LPTx-\CLPSx,\CLP,\Popeetala,\NastVam}). Here
we briefly review this technique, and then extend it to the
consideration of the dilaton and axion in IIB supergravity.

\subsec{The exact form of the internal metric}

The starting point is the encoding of the gauge fields of the
dimensionally reduced theory into the metric and Killing vectors of
the parent theory.  While this technique had always been a staple of
dimensional reduction at the linearized order, it was argued in \UCAK\
that when properly stated, such encoding of gauge fields must be exact
to all orders in fields.  The argument was based upon how the gauge
symmetries in the reduced theory must be related to the diffeomorphism
invariance of the parent theory, and that this relationship would be
spoilt if the linearized Ansatz were not, in fact, exact.

To be more specific, consider a theory in $D$-dimensions that is
reduced to $d$-dimensions on a manifold, ${\cal M}$, with isometries
represented by Killing fields: $K^{\Omega \, p}$, where $\Omega$ indexes the
Killing fields and $p$ is the vector index on ${\cal M}$.  Decompose
the $D$-dimensional vielbein according to:
\eqn\vielb{e_M{}^A \eql  \left(\matrix{ e_\mu{}^\alpha & e_\mu{}^a \cr
e_m{}^\alpha & e_m{}^a \cr}\right) \ ,}
where the upper left corner represents the $d$-dimensional space-time
and the bottom right represents the $(D-d)$-dimensioanal
manifold ${\cal M}$.  The claim of \UCAK\ is that
\eqn\Axact{e_\mu{}^a \eql  A_\mu^\Omega \,K^{\Omega \, p}  \,e_p{}^a }
is the exact consistent truncation Ansatz for the gauge fields
to all orders.

It was observed in \MetAns\ that when the foregoing is combined with
the supersymmetry transformations of the gauge fields, one could
obtain the exact Ansatz for the internal metric $g_{pq}$.  To
illustrate this we consider the $S_5$ reduction of the IIB
supergravity to five dimensions, but we stress that the argument is
very general.  Consider the gravitino terms in the supersymmetry
variations of the five-dimensional gauge vector fields
\refs{\GRW,\PPvN}:
\eqn\Asusyvars{\delta A^{IJ}_\mu \eql  2 i \,\widetilde \cV_{IJab}\,
\overline {\widehat{\epsilon} }^{\, a} \,
\widehat{\psi}_\mu^b \,+\, \dots \, \ . }
In this equation, $I,J =1,\dots,6$, and $A^{IJ} = -A^{JI}$
represent the  $SO(6)$ gauge fields.  The hats $~\widehat {}~$
have been introduced to distinguish five-dimensional fields
from their ten-dimensional antecedents.

Using \Axact\ the corresponding ten-dimensional variation
of the vielbein gives:
\eqn\Esusyvars{\delta A^{IJ}_\mu \,K^{IJ \, p} \eql
(\delta e_\mu{}^a)\,e_a{}^p \,+\, e_\mu{}^a\, (\delta e_a{}^p)
\eql  - 2 \kappa\, {\rm Im}\,(
 {\overline \epsilon}   \gamma^a \psi_\mu ) \,+\,\dots \, \ . }

One now recalls that the dimensional reduction to the
standard Einstein action and Rarita-Schwinger actions
in $d$ dimensions requires a proliferation of ``warp'' factors.
In particular, one has:
\eqn\warps{ e_\mu{}^\alpha \eql  \Delta^{-{1 \over d-2}}\,
\widehat e_\mu{}^\alpha\, ; \quad \psi_\mu \eql  
 \Delta^{-{1 \over 2(d-2)}}\,
\widehat \psi_\mu  \, ; \quad   \epsilon \eql
 \Delta^{-{1 \over 2(d-2)}}\,
\widehat \epsilon \ ,}
where the hats refer to $d$-dimensional quantities, and the
warp-factor is given by:
\eqn\Deltadefn{\Delta \,\equiv\,{\rm  det}\,( e_p{}^a \,
\eop_b{}^p)  \eql  \sqrt{{\rm det}( g_{mp}\,\gop^{pq})} \ . }
The inverse frame $\displaystyle \eop_b{}^p$ and the inverse metric
$\displaystyle \gop^{pq}$ are those of the ``round,'' maximally
supersymmetric  background on ${\cal M}$.  The metric warp-factor is
introduced so that the $D$-dimensional Einstein action, along
with its factors of $\sqrt{g}$ reduce to the $d$-dimensional
Einstein action.  The Rarita-Schwinger field is rescaled for
the same reason, and the supersymmetry parameter is rescaled
so that one gets the canonical form for the supersymmetry
variations of $\widehat \psi_\mu$ and $\widehat e_\mu{}^\alpha$ in
$d$-dimensions.  The effect of all this warping is that
\Esusyvars\ becomes:
\eqn\Esusywarp{\delta A^{IJ}_\mu \,K^{IJ \, p}   \eql  - 2 \kappa\,
{e_a}{}^p\, \Delta^{-{1 \over d-2}}\,{\rm Im}( {\overline{
\widehat\epsilon}} \gamma^a
\widehat  \psi_\mu ) \,+\,\dots \, \ . }

One now compares \Asusyvars\ with \Esusywarp.  The internal
indices of the former, arise through the labelling of
Killing spinors $\cM$.  For the five-dimensional supergravity
this labelling is ambiguous up to a $USp(8)$ transformation,%
\foot{Recently, equations that determine those $USp(8)$ transformations
were obtained in \NastVam.}
but such internal local symmetry transformations can be eliminated
by squaring and contracting with the $USp(8)$ symplectic
form: $\Omega^{ab}$.  The result is the squaring and
contraction of the inverse-vielbein ${e_a}^p\,$ in \Esusywarp\
yields the inverse metric, and one obtains (putting $d=5$):
\eqn\metanswer{\Delta^{-{2 \over 3}}\,\widehat g^{pq} \eql 
 {1 \over a^2}\, K^{IJ\, p}\,K^{KL\, q}\, \widetilde
\cV_{IJab}\,\widetilde
\cV_{KLcd}\, \Omega^{ac}\,\Omega^{bd} \ ,}
where the constant, $a$, depends upon the normalization of
the Killing vectors.

This argument was performed in great detail in \MetAns\ for
the $S_7$ reduction of eleven-dimensional supergravity.  The
foregoing argument was also made to arrive at the expression
for the general IIB compactification metric given in
\KPW.  The details of the local $USp(8)$ structure in
ten dimensions were not explicitly checked for the result in
\KPW, and so for that reason we referred to it as a ``conjecture,''
however, a more precise statement of that result would have
been: If the truncation is consistent, then the internal
metric must be given by \metanswer.

\subsec{The exact form for the dilaton}

In the same spirit as in \KPW, we will now  conjecture an
exact form for the ten-dimensional dilaton.

The starting point is now the encoding of the five-dimensional
tensor gauge fields in the ten-dimensional, two-form
gauge potentials $A^\alpha_{MN}$.  At the linearized order one has:
\eqn\BfieldAns{A^\alpha_{\mu \nu}\eql
{\widehat B}^{I \alpha}_{\mu \nu} \,x^I\ ,}
where $\alpha =1,2$ now denotes an $SL(2,\IR)$ index, $I$ is the
$SO(6)$ vector index, and $x^I$ are the cartesian coordinates
of the $5$-sphere: $\sum_I (x^I)^2 =1$.

It seems plausible, based on the gauge symmetries, the minimal
couplings, and mixings with the gauge fields, that this linearized
Ansatz is exact to all orders.  Rather than prove this in detail, we
shall assume that it is true and derive the dilaton Ansatz.  The body
of this paper then represents a highly non-trivial test of this
assumption.

One proceeds exactly as in the previous subsection, except that one
compares specific gravitino terms in the ten-dimensional and
five-dimensional supersymmetry variations of the two-form field
strengths:
\eqn\Atwovars{\eqalign{\delta A^\alpha_{\mu \nu}\eql  & 4i\, V^\alpha_+
\,
 {\bar \epsilon}   \gamma_{[\mu} \psi^*_{\nu ]} ) \,+\, 4i\,
V^\alpha_- \, {\bar \epsilon}^*   \gamma_{[\mu} \psi_{\nu]} )
\,+\, \dots \cr \eql
& \Delta^{-{2 \over3}} \,\Big( 4i\, V^\alpha_+ \, \hat {\bar \epsilon}
 \gamma_{[\mu} \hat \psi^*_{\nu ]} ) \,+\, 4i\,  V^\alpha_- \, \hat
{\bar
\epsilon}^*   \gamma_{[\mu}\hat \psi_{\nu]} ) \Big) \,+\, \dots \ ,}}
and
\eqn\Btwovars{\delta \widehat B^{I \alpha}_{\mu \nu}\eql  2 i g\,
\epsilon^{\alpha \beta}\, \cV_{I \beta a b} \,
{\overline{\widehat \psi}}{}^{\, a}_{[\mu}  \gamma_{ \mu]} \epsilon^b
\,+\, \dots \ .}
Again one can remove the local $USp(8)$ tansformations by
squaring and contracting to obtain:
\eqn\SAns{\Delta^{-{4 \over3}}\, (S\,S^T)^{\alpha \beta} \eql
{\rm const}\times
  \,\epsilon^{\alpha \gamma} \epsilon^{\beta \delta}\,
\cV_{I \gamma}{}^{ a b} \,\cV_{J \delta}{}^{ c d} 
\,x^I x^J\, \Omega_{ac}\,
\Omega_{bd} \ ,}
where ${S^\alpha}_\beta$ is the IIB dilaton/axion matrix
written in the $SL(2,\IR)$ basis, with the local $U(1)
=O(2)$ acting from the right.  (Note that $V^\alpha_\pm$
of \JSIIB\ is generally written in the $SU(1,1)$ basis.)

One obvious consistency check is that the value of $\Delta$
obtained from taking the determinants of both sides of
\SAns\ and of \metanswer\ agree.  This was indeed confirmed in
section 3.2.

One can choose a gauge in which $S \in SL(2,\IR)$ is symmetric, and so
one can take the square root of \SAns, and extract $S$.  One should
also note that if the $E_{6(6)}$ matrix $\cV$ of five-dimensional
scalars is, in fact, in the subgroup $SL(6,\IR) \times SL(2,\IR)$ used
in \GRW, then the IIB dilaton and axion are indeed precisely
described by the $SL(2,\IR)$ factor of this $SL(6,\IR) \times
SL(2,\IR)$.  The formula \SAns\ is consistent with this special
case.  However, for general the $E_{6(6)}$ matrix $\cV$, the
five-dimensional and ten-dimensional $SL(2,\IR)$ factors have a highly
non-trivial relationship.

In terms of the AdS/CFT correspondence, this last statement
means that on the Coulomb branch of the $\cN =4$ Yang-Mills
theory, the gauge coupling and theta angle are constant, and
are represented by the $SL(2,\IR)$ of the five-dimensional
theory.  However, if fermion masses are turned on in the
gauge theory, then \SAns\ tells us exactly how the running of
the gauge coupling and  axion are determined entirely in terms of the
running of all the masses of the Yang-Mills scalar and fermion fields.
The $SL(2,R)$ matrix of the five-dimensional theory is a
global symmetry of the potential and is presumably broken to
$SL(2,\ZZ)$ in the quantum theory: it represents the symmetry
of even the perturbed theory under the  $SL(2,\ZZ)$ action
on the $\cN=4$ coupling.  The actual physical coupling,
represented by the ten-dimensional dilaton and axion, are
non-trivial functions of this $\cN=4$ coupling, the masses of
the fields and the scale.  This relationship is given in
large $N$ theories by \SAns.

\listrefs
\vfill
\eject
\end